\documentclass[12pt]{article}
\usepackage{amsmath,amssymb,amsfonts,amsthm}
\title{A theoretical scheme for generation of Gazeau-Klauder coherent states via
         intensity-dependent degenerate Raman interaction}

 \author{F. Yadollahi, M. K. Tavassoly
\\
\footnotesize{Atomic and Molecular Group, Faculty  of Physics,
Yazd University, Yazd, Iran}
\\ \footnotesize{e-mail: mktavassoly@yazduni.ac.ir  } }
\begin{document}

 \date{\today}


 \newcommand{\I}{\mathbb{I}}
 \newcommand{\norm}[1]{\left\Vert#1\right\Vert}
 \newcommand{\abs}[1]{\left\vert#1\right\vert}
 \newcommand{\set}[1]{\left\{#1\right\}}
 \newcommand{\R}{\mathbb R}
 \newcommand{\C}{\mathbb C}
 \newcommand{\DD}{\mathbb D}
 \newcommand{\eps}{\varepsilon}
 \newcommand{\To}{\longrightarrow}
 \newcommand{\BX}{\mathbf{B}(X)}
 \newcommand{\HH}{\mathfrak{H}}
 \newcommand{\D}{\mathcal{D}}
 \newcommand{\N}{\mathcal{N}}
 \newcommand{\W}{\mathcal{W}}
 \newcommand{\RR}{\mathbb{\mathcal}{R}}
 \newcommand{\HD}{\hat{\mathcal{H}}}
  \maketitle

 \begin{abstract}

   A theoretical scheme is presented for generating Gazeau-Klauder coherent states (GKCSs)
   via the generalization of degenerate Raman interaction with coupling  constant to intensity-dependent coupling.
   Firstly, we prove that in the intensity-dependent degenerate Raman interaction,
   under particular conditions, the modified effective Hamiltonian
   can be used instead of Hamiltonian in the interaction picture, for describing the atom-field interaction.
   We suppose that the cavity field is
   initially prepared in a nonlinear CS, which is not
   temporally stable. As we will observe, after the occurrence of the interaction between atom and field,
   the generated state involves a superposition of GKCSs which are temporally stable and initial nonlinear CS.
   Under specific conditions which may be prepared, the generated state just includes GKCS.
   So, in this way we produced the GKCS, successfully.
 \end{abstract}






\section{Introduction}\label{sec-intro}
      Coherent states (CSs) defined as the right eigenstates of the harmonic oscillator annihilation operator,
      i.e., $a|\alpha\rangle=\alpha |\alpha\rangle$ play an important role in quantum optics and modern physics
      \cite{Ali}. Along the generalization of these states, nonlinear CSs \cite{Vogel} or $f$-CSs \cite{manko}
      have been introduced. According to this formalism $f$-deformed annihilation and creation operators,
      respectively defined as $A=a f(n)$ and $A^\dag= f^\dag (n) a^\dag$ where $a,a^\dag$ and $n=a^\dag a$
      are bosonic annihilation, creation and number operators, respectively. The intensity-dependent
      function $f(n)$ is responsible for the nonlinearity of the states. Nonlinear CSs $|z,f\rangle$
      are then defined as the right eigenstates of $f$-deformed
      annihilation operator,
  \begin{equation}\label{vizhe megh nonlin}
      A|z,f\rangle=z|z,f\rangle.
  \end{equation}
      The Fock space representation of these states is explicitly given by
  \begin{eqnarray}\label{bast nonlin}
      &|z,f\rangle&=\sum_{n=0}^{\infty}C_{n}|n\rangle ,\qquad C_{n}=\mathbf{\mathcal{N}}(|z|^2)\;\frac{z^n}
      {\sqrt{n!}\;[f(n)]!},\qquad\nonumber\\
      &\mathbf{\mathcal{N}}(|z|^2)&=\left (\sum_{n=0}^{\infty} \frac{|z|^{2n}}{n!
      \,([f(n)]!)^2}\right)^{-1/2},\qquad z\in \mathbb{C}
  \end{eqnarray}
      where $[f(n)]!=f(1)f(2)...f(n)$ and by convention $[f(0)]!\doteq 1$. Note that we have confined
      ourselves to the special case of real valued function $f(n)$.

      Roknizadeh et al derived a Hamiltonian associated to a nonlinear system based on action identity
      requirement of nonlinear coherent states as $ H=A^\dag A=n f^2 (n)$ \cite{rokni-tav-8111}.
      So, the eigenvalue equation for any one dimensional physical system with known discrete eigenvalues
      may be given by $H|n\rangle=e_{n}|n\rangle=n f^2(n)|n\rangle$, where
      $0=e_{0}<e_{1}<e_{2}<...<e_{n}<e_{n+1}<...\;$. Accordingly, one simply
      has $f(n)=\sqrt{{e_{n}}/{n}}$ associated to solvable quantum systems.
      In this way the nonlinear CSs may be deduced corresponding to any solvable
      quantum system as \cite{Honarasa}:
  \begin{equation}\label{bast nonlinear En}
      |z,e_{n}\rangle=\mathbf{\mathcal{N}}(|z|^2)\sum_{n=0}^{\infty} \frac{z^n}{\sqrt{[e_{n}]!}}|n\rangle.
  \end{equation}
      But it is remarkable that these states are not temporally stable like the original nonlinear CSs
      in (\ref{bast nonlin}).
      Moreover, an analytical representation of Gazeau-Klauder CSs (GKCSs) corresponding to
      any Hamiltonian with discrete (nondegenerate) eigenvalues is defined as
       \cite{Gazeau, rokni-tav-042110, Kinani}
  \begin{equation}\label{bast GK z,alpha}
       |z,\alpha\rangle=\mathbf{\mathcal{N}}(|z|^2)\sum_{n=0}^{\infty}\frac{z^n
        \,e^{-i\alpha e_{n}}}{\sqrt{\rho(n)}}|n\rangle,\;\;\;\;\;\;\;z\in
        \mathbf{\mathbb{C}},\;\;\;\;\;\;\;\alpha\in\mathbf{\mathbb{R}},
  \end{equation}
        where $\mathbf{\mathcal{N}}(|z|^2)$ is a normalization constant.
        These states satisfy the following requirements: (i) continuity of label,
        (ii) resolution of the identity, (iii) temporal stability and (iv) action identity.
        The last condition requires $\rho(n)=[e_{n}]!$. So the state in (\ref{bast GK z,alpha}) can be written as
  \begin{equation}\label{final bast GK z,alpha}
        |z,\alpha\rangle=\mathbf{\mathcal{N}}(|z|^2)\sum_{n=0}^{\infty}\frac{z^n
        \,e^{-i\alpha e_{n}}}{\sqrt{[e_{n}]!}}|n\rangle.
  \end{equation}
       It is also established in \cite{rokni-tav-042110} that GKCSs
       are a class of nonlinear CSs with nonlinearity function $f_{GK}(\alpha,n)=
       \sqrt{\frac{e_{n}}{n}}\;e^{i\alpha (e_{n}-e_{n-1})}$, which are temporally stable, i.e.,
  \begin{equation}\label{temporal stable}
        e^{-iHt}|z,\alpha\rangle=|z,\alpha '\rangle,\;\;\;\;\;\;\;\;\;\;\alpha '\equiv\alpha +t.
  \end{equation}
       Along this work, Gazeau-Klauder squeezed states associated to
       solvable quantum systems have been introduced by one of us
       \cite{gkss}.

       On the other hand, recently there has been much interest in the superpositions
       of CSs \cite{Abbasi}. Due to the quantum interference between the coherent
       components, such superposition states may exhibit various nonclassical
       properties such as squeezing and sub-Poissonian statics. Nonclassical
       states of light are central to quantum optics. Their importance comes
       from potential applications on advanced optics, as teleportation \cite{1},
       quantum computation \cite{2}, quantum communication \cite{3}, quantum
       cryptography \cite{4}, quantum lithography \cite{5}, etc. A number of
       schemes have been proposed for generating such states \cite{strongly squeezed}.
       Among them, a method has been presented for generating superpositions of CSs of
       a cavity field via degenerate Raman interaction \cite{generation1,generation2}.
       In this paper, we have followed the same approach and suggested a theoretical scheme
       for generating the  GKCSs which are temporally stable, via the
       intensity-dependent degenerate Raman interaction. It is worth
       to mention that, we have used the intensity-dependent modified
       effective Hamiltonian which describes appropriately the dynamics of the interaction between atom and field.
       This is while, to the best of our knowledge,
       no scheme for generation of GKCSs may be found in the earlier literature.

 \section{Degenerate Raman interaction and modified effective Hamiltonian }
      The degenerate Raman interaction describes the interaction between a degenerate
      $\Lambda$-type three-level atom (as shown in Fig. 1) and a
      single-mode radiation field characterized by bosonic operators $a, a^\dag$.
      Various properties of this system have been previously discussed
      \cite{interact Hamilt1,interact Hamilt2}. The Hamiltonian for such
      a system in the interaction picture is given by:
   \begin{eqnarray}\label{interaction Hamiltonian}
       H_{I} = g_{1} (a^\dag |g\rangle \langle i| e^{-i\triangle t}+ a
       |i\rangle \langle g| e^{i\triangle t})
       + g_{2} (a^\dag |e\rangle \langle i| e^{-i\triangle t}+ a |i\rangle \langle e| e^{i\triangle t}),
  \end{eqnarray}
        where we have set $\hbar\equiv1$, $|e\rangle$ and $|g\rangle$ are the two
        degenerate lower states and $|i\rangle$ is the upper state of the atom,
        the coupling constant of the transition between $|i\rangle$ and $|g\rangle
        (|e\rangle)$ with the cavity field is denoted by  $g_{1}(g_{2})$,
        $\triangle=(\omega _{i}-\omega _{0})-\omega _{f}$ is detuning,
        where $\omega _{f}$ is the cavity field frequency, $\omega _{0}$
        is the energy of the two lower states and $\omega _{i}$ is the
        energy for the upper level of the atom.
          On the other side, to describe the degenerate Raman interaction, an effective Hamiltonian was derived by Agarwal in \cite{Agarwal},
          when the atomic transition between the upper
          and lower levels is far from the frequency of the field mode.
          Indeed, after adiabatically eliminating the upper level, the author introduced the effective Hamiltonian
          as:
   \begin{eqnarray}\label{effective Hamiltonian1}
        H_{e} &=& -\lambda\, a^\dag a\, (|g\rangle \langle e| + |e\rangle \langle g|),
   \end{eqnarray}
        where $\lambda=g_{1}g_{2}/\triangle$ indicates the effective atom-field coupling constant.
        In fact, in the effective Hamiltonian approach, the two lower levels are coupled with a
        single-mode field through a virtual upper level. As a result,
        the degenerate Raman interaction is reduced to an effective two-level system.
        Also, the effective Hamiltonian given in (\ref{effective Hamiltonian1}) leads to the elimination of time-dependent
        phase factors in (\ref {interaction Hamiltonian}). These factors are related
        to the ac Stark shifts of energy levels \cite{ac stark}.
        The Hamiltonian that represents the Stark shifts of energy levels is given by:
  \begin{eqnarray}\label{ac stark}
       H_{s} = - a^\dag a\, (\lambda_{1}|g\rangle \langle g|+ \lambda_{2}|e\rangle \langle e|),
  \end{eqnarray}

       where $\lambda_{1}=g_{1}^2/\Delta \,(\lambda_{2}=g_{2}^2/\Delta)$ describes the Stark shift of the state $|g\rangle \,(|e\rangle)$.
       Hence, one may modify the effective Hamiltonian given in (\ref{effective Hamiltonian1}),
        so that it includes the ac Stark effects represented in (\ref{ac stark}). In this way, for the
        modified effective Hamiltonian, one has then \cite{modified}
  \begin{eqnarray}\label{sum Hamiltonian}
        H_{eff}= H_{e}+H_{s}.
  \end{eqnarray}
      Inserting $H_{e}$ and $H_{s}$ respectively from (\ref{effective Hamiltonian1}) and (\ref{ac stark})
      in (\ref{sum Hamiltonian}), the modified effective Hamiltonian can be written as follows:
 \begin{eqnarray}\label{modified effective Hamiltonian}
       H_{eff}= -\lambda \,a^\dag a\, (|g\rangle \langle e|+
       |e\rangle \langle g|)
       -a^\dag a\,(\lambda_{1}|g\rangle \langle g|+\lambda_{2}|e\rangle \langle e|).
  \end{eqnarray}
       It is found that, the modified effective
       Hamiltonian $H_{eff}$ is more suitable than the well-known effective
       Hamiltonian $H_{e}$ in studying the time evolution of the systems.
       Xu and Zhang in \cite{modified} showed that, in the degenerate Raman interaction,
       far off-resonant (enough large value of $\Delta$), the
       modified effective Hamiltonian has been given in (\ref{modified effective Hamiltonian}), can be applied
       instead of the Hamiltonian in the interaction picture in (\ref{interaction Hamiltonian}).
       In fact, starting from a unique state vector at $t=0$,
       both of the Hamiltonians lead to the same state vector of atom-field system at any time $t$.
       Altogether, several other mathematical
       physics methods were suggested to obtain the modified effective Hamiltonian in \cite{numercial1,numercial2}.

   \section{Generalizing the degenerate Raman interaction to intensity-dependent interactions}

       Generalization of the single-mode to two-mode degenerate Raman process has been
       recently done in \cite{ShiBiao Zheng}.
       But, to achieve our aim of the paper, we should generalize the single-mode degenerate Raman interaction in a rather different way.
       Indeed, following the path of Knight in \cite{Knight} and specifically Zheng in
       \cite{Zheng},
       we replace the bosonic operators $a, a^\dag$ in (\ref{modified effective Hamiltonian})
       with $A=a\,f(n), A^\dag=f(n)a^\dag$, i.e., the generalized $f$-deformed ladder operator.
       This type of generalization may be arisen from the same procedure of degenerate Raman interaction, briefly
       illustrated in the previous section, with the difference that the interaction between
       the same $\Lambda$-type atom and a single-mode radiation field
       is now considered to be intensity-dependent.
       Henceforth, analogously to (\ref{interaction Hamiltonian}) one may suggest the following form for the Hamiltonian of such
      a system in the interaction picture
   \begin{eqnarray}\label{tamim interaction Hamiltonian}
       \mathbf{\mathcal{H}}_{I} = g_{1} (A^\dag |g\rangle \langle i| e^{-i\triangle t}+
       A |i\rangle \langle g| e^{i\triangle t})
       + g_{2} (A^\dag |e\rangle \langle i| e^{-i\triangle t}+ A |i\rangle \langle e| e^{i\triangle t}).
  \end{eqnarray}
       In the same way, analogously to (\ref{modified effective Hamiltonian}),
       for the intensity-dependent modified effective Hamiltonian we suggest
  \begin{eqnarray}\label{tamim effective Hamiltonian}
       \mathbf{\mathcal{H}}_{eff}= - \lambda \,A^\dag A\, (|g\rangle \langle e|+
       |e\rangle \langle g|)
       -A^\dag A\,(\lambda_{1}|g\rangle \langle g|+\lambda_{2}|e\rangle \langle e|).
  \end{eqnarray}
       Clearly, inserting $f(n)=1$ in (\ref{tamim interaction Hamiltonian}) and
       (\ref{tamim effective Hamiltonian}) recovers the usual
       degenerate Raman interaction (intensity independent coupling) introduced in
       (\ref{interaction Hamiltonian}) and (\ref{modified effective Hamiltonian}), respectively.
       This type of generalization (of degenerate Raman interaction) to intensity-dependent interaction
       can be frequently found in the literature, for instance, whenever one imposes nonlinearity on the standard JCM \cite{five ref}.

    Before we proceed, to establish the above formal approach more explicitly,
    the following discussion may be offered.
    As mentioned, when $\Delta$ is enough large,
    the equivalence of the introduced Hamiltonian in (\ref{interaction Hamiltonian})
    and (\ref{modified effective Hamiltonian}) are demonstrated in \cite{modified}.
    Our aim at this stage is to prove that, in the nonlinear degenerate Raman
    interaction, under particular conditions, the
    nonlinear modified effective Hamiltonian in (\ref{tamim effective Hamiltonian}) can be
    applied appropriately instead of the nonlinear Hamiltonian in the interaction
    picture introduced in (\ref{tamim interaction Hamiltonian}).
    To investigate the validity of our proposal, we will obtain the time evolution of the state vector of the atom-field system
    by using  (\ref{tamim interaction Hamiltonian})
    and  (\ref{tamim effective Hamiltonian}), and continue with showing that
    for enough large value of $\Delta$, these two state vectors are equal.
    For this purpose, we suppose that the atom is initially in a superposition of $|g\rangle$ and $|e\rangle$:
 \begin{eqnarray}\label{atom state t=0}
       |\Phi_{a}(t=0)\rangle=C_{g}(0) |g\rangle+C_{e}(0) |e\rangle,
 \end{eqnarray}
    and the field is in the arbitrary state:
 \begin{eqnarray}\label{Hamdus}
   |\Phi_{F}(t=0)\rangle=\sum_{n=\circ}^{\infty} q_{n}|n\rangle,
 \end{eqnarray}
   where $q_{n}$ determines the initial state of the field. The state of the atom-field system at $t=0$ is thus given by:
 \begin{eqnarray}\label{Halate system t=0}
  |\Phi_{s}(t=0)\rangle=\sum_{n=0}^{\infty} q_{n}\;( C_{g}(0) |g,n\rangle+C_{e}(0) |e,n\rangle).
 \end{eqnarray}
   The time evolution of the state of the atom-field system, governed
   by $\mathbf{\mathcal{H}}_{I}$ introduced in (\ref{tamim interaction Hamiltonian}), is the solution of:
 \begin{eqnarray}\label{schrodinger equation}
      i\frac{\partial}{\partial t}|\Phi_{s,I}(t)\rangle=
      \mathbf{\mathcal{H}}_{I} |\Phi_{s,I}(t)\rangle,
  \end{eqnarray}
    where $|\Phi_{s,I}(t)\rangle$, denotes the state of the atom-field system  after time
    $t$.
  The following general solution may be considered as:
  \begin{eqnarray}\label{Halate system t barhamkonesh}
    |\Phi_{s,I}(t)\rangle=\sum_{n=\circ}^{\infty} q_{n}\;\left[ C_{g}^{n}(t) |g,n\rangle+C_{i}^{n}(t) |i,n-1\rangle+C_{e}^{n}(t) |e,n\rangle
    \right].
  \end{eqnarray}
  The problem is now determining the time dependent coefficients which
  may be obtained by lengthy but straightforward calculations with the following final results:
 \begin{eqnarray}\label{zarayeb Int11}
    C_{g}^{n}(t) = A_{1}^{n}(t)\;C_{g}(0)+ A_{2}^{n}(t)\;C_{e}(0),
 \end{eqnarray}
 \begin{eqnarray}\label{zarayeb Int22}
     C_{i}^{n}(t) = B_{1}^{n}(t)\;C_{g}(0)+ B_{2}^{n}(t)\;C_{e}(0),
 \end{eqnarray}
 \begin{eqnarray}\label{zarayeb Int33}
  C_{e}^{n}(t) = A_{2}^{n}(t)\;C_{g}(0)+ A_{3}^{n}(t)\;C_{e}(0),
 \end{eqnarray}
    where we have set
 \begin{eqnarray}\label{zarayeb Int1}
   A_{1}^{n}(t)=e^{-i\Delta t/2} \left(\frac{g_{2}^2}{G}\;e^{i\Delta t/2}+\frac{g_{1}^2}{G} \;
   \cos \Lambda_{n}t+i \frac{\Delta g_{1}^2}{2 \Lambda_{n}G}\;\sin \Lambda_{n}t \right),
 \end{eqnarray}
 \begin{eqnarray}\label{zarayeb Int2}
   A_{2}^{n}(t)=e^{-i\Delta t/2} \left(-\frac{g_{1}g_{2}}{G}\;e^{i\Delta t/2}+\frac{g_{1}g_{2}}{G}
   \;\cos \Lambda_{n}t+i \frac{\Delta g_{1}g_{2}}{2 \Lambda_{n}G}\;\sin \Lambda_{n}t \right),
 \end{eqnarray}
 \begin{eqnarray}\label{zarayeb Int3}
   A_{3}^{n}(t)=e^{-i\Delta t/2} \left(\frac{g_{1}^2}{G}\;e^{i\Delta t/2}+\frac{g_{2}^2}{G} \;\cos
   \Lambda_{n}t+i \frac{\Delta g_{2}^2}{2 \Lambda_{n}G}\;\sin \Lambda_{n}t \right),
 \end{eqnarray}
 \begin{eqnarray}\label{zarayeb Int4}
   B_{1}^{n}(t)=e^{i\Delta t/2}\left (-i\frac{g_{1}\sqrt{n} f(n)}{\Lambda_{n}}\;\sin \Lambda_{n}t\right),
 \end{eqnarray}
 \begin{eqnarray}\label{zarayeb Int5}
   B_{2}^{n}(t)=e^{i\Delta t/2}\left (-i\frac{g_{2}\sqrt{n}f(n)}{\Lambda_{n}}\;\sin \Lambda_{n}t\right),
 \end{eqnarray}
  with  the following  definitions:
 \begin{eqnarray}\label{lambda n}
  \Lambda_{n} \doteq \sqrt{nf^2(n) G+\frac{\Delta^2}{4}},
 \end{eqnarray}
 \begin{eqnarray}\label{G}
   G  \doteq g_{1}^2+g_{2}^2.
 \end{eqnarray}
   On the other side, if we use $ \mathbf{\mathcal{H}}_{eff}$ introduced in
   (\ref{tamim effective Hamiltonian}) for obtaining the time evolution of the state of the atom-field,
   we have to solve the equation
 \begin{eqnarray}\label{schrodinger equation}
  i\frac{\partial}{\partial t}|\Phi_{s,eff}(t)\rangle =
  \mathbf{\mathcal{H}}_{eff} |\Phi_{s,eff}(t)\rangle,
 \end{eqnarray}
   where $|\Phi_{s,eff}(t)\rangle$ denotes the state of the atom-field system after time $t$.
   The general solution may be given by
 \begin{eqnarray}\label{Halate system t moaser}
   |\Phi_{s,eff}(t)\rangle=\sum_{n=0}^{\infty} q_{n}\;\left[ C_{g}^{n}(t) |g,n\rangle+C_{e}^{n}(t) |e,n\rangle \right]
 \end{eqnarray}
 with
  \begin{eqnarray}\label{zarayeb moaser11}
     C_{g}^{n}(t) = D_{1}^{n}(t)C_{g}(0)+ D_{2}^{n}(t)C_{e}(0),
  \end{eqnarray}
  \begin{eqnarray}\label{zarayeb moaser22}
   C_{e}^{n}(t) = D_{2}^{n}(t)C_{g}(0)+ D_{3}^{n}(t)C_{e}(0).
  \end{eqnarray}
  Here we have set
 \begin{eqnarray}\label{zarayeb moaser1}
  D_{1}^{n}(t) =\frac{g_{2}^2}{G} +\frac{g_{1}^2}{G} \exp(\frac{i n f^2(n)G}{\Delta}\;t) ,
 \end{eqnarray}
 \begin{eqnarray}\label{zarayeb moaser2}
   D_{2}^{n}(t) = -\frac{g_{1}g_{2}}{G}+ \frac{g_{1}g_{2}}{G}\exp(\frac{i nf^2(n) G}{\Delta}),
 \end{eqnarray}
 \begin{eqnarray}\label{zarayeb moaser3}
  D_{3}^{n}(t) =\frac{g_{1}^2}{G} +\frac{g_{2}^2}{G} \exp(\frac{i n f^2(n)G}{\Delta}\;t) .
 \end{eqnarray}
 When $ 4\bar{n} f^2(\bar{n})\ll\Delta^2/G$ namely $\Delta$ be
 enough large and $(\bar{n} f^2(\bar{n}))^2 G^2 t/\delta^3 \ll \pi$
 which means that evolving time is not too long, one has:
 \begin{eqnarray}\label{delta bozorg natayej1}
  \frac{\Lambda_{n}}{\Delta}\rightarrow \frac{1}{2},
 \end{eqnarray}
 \begin{eqnarray}\label{delta bozorg natayej1}
   \Lambda_{n}-\frac{\Delta}{2} \rightarrow \frac{n f^2(n) G}{\Delta}.
 \end{eqnarray}
 Therefore, by using the two above approximations we have:
 \begin{eqnarray}\label{tabdile zarayeb1}
   A_{k}^{n}(t)\rightarrow  D_{k}^{n}(t),\qquad k=1,2,3
 \end{eqnarray}
 \begin{eqnarray}\label{tabdile zarayeb2}
  B_{i}^{n}(t)\rightarrow \circ  ,\qquad i=1,2,
 \end{eqnarray}
 and consequently,
 \begin{eqnarray}\label{convertation}
 |\Phi_{s,I}(t)\rangle\rightarrow |\Phi_{s,eff}(t)\rangle.
 \end{eqnarray}
 As it is observed, we established that in the nonlinear degenerate
 Raman interaction, whenever $\Delta$ be enough large, the state
 vectors given in (\ref{Halate system t barhamkonesh}) and
 (\ref{Halate system t moaser}) will be equal.
 Therefore, in the nonlinear degenerate Raman interaction, when $\Delta$ be enough large,
 the nonlinear modified effective Hamiltonian is equivalent to the nonlinear
 Hamiltonian in the interaction picture.

       \section{Generating of GKCSs via the intensity-dependent degenerate Raman interaction}
       Now,  in the remainder of the paper, we want to present our theoretical scheme for generating
       temporal stable GKCSs via the introduced intensity-dependent degenerate Raman interaction, while the atom-field interaction is described by the modified effective Hamiltonian given in (\ref{tamim effective Hamiltonian}).
       It is worth to notice that, without loss of generality, $g_{1}=g_{2}=g$ is assumed
       in (\ref{tamim effective Hamiltonian}) for continuing the calculational works, thus one has $\lambda_{1}=\lambda_{2}=\lambda = g^2/\triangle$.
       So, one arrives at the modified effective Hamiltonian as follows:
  \begin{eqnarray}\label{tamim modified effective Hamiltonian}
       \mathbf{\mathcal{H}}_{eff}= - \lambda A^\dag A\, (|g\rangle \langle e|+
       |e\rangle \langle g|+|g\rangle \langle g|+|e\rangle \langle e|).
  \end{eqnarray}
       Using the expression of $A, A^\dag$,  it may be readily seen that the effective Hamiltonian in
       (\ref{tamim modified effective Hamiltonian}) can be explicitly re-written as:
       \begin{eqnarray}\label{tamime effective Hamiltonian,rewrite}
       \mathbf{\mathcal{H}}_{eff}= - \lambda\, f^2(n)\, a^\dag a\, (|g\rangle \langle e|+
       |e\rangle \langle g|+|g\rangle \langle g|+|e\rangle \langle e|).
  \end{eqnarray}
       Inserting $\lambda_1=\lambda_2= \lambda $ in (\ref{modified effective Hamiltonian}) and then comparing
       the recasted relation with (\ref{tamime effective Hamiltonian,rewrite}),
       the above development can be, in a sense, considered simply as
       generalizing $\lambda$ to $\lambda f^2(n)$. This result
       may be viewed as intensity-dependent atom-field coupling.

       Now, suppose that the cavity field is initially prepared in the nonlinear CS introduced in
      (\ref{bast nonlin}).  Also, the flux of the atoms injected into the cavity is such a low
       that there exists at most one atom at a time inside the cavity. If the first
       atom be initially in a superposition of $|e\rangle$ and $|g\rangle$, such that
 \begin{eqnarray}\label{ first atom}
       |\Psi_{a}^{(1)}(0)\rangle=\frac{1}{\sqrt{1+|\varepsilon_{1}|^2}} (|e\rangle+\varepsilon_{1}|g\rangle),
  \end{eqnarray}
       the state of the atom-field system at $t=0$ is thus given by:
  \begin{eqnarray}\label{first atom+field at t=0}
      |\Psi_{s}^{(1)}(t=0)\rangle=\frac{1}{\sqrt{1+|\varepsilon_{1}|^2}}
      \sum_{n=0} ^{\infty} C_{n} (|e,n\rangle+\varepsilon_{1}|g,n\rangle).
 \end{eqnarray}
      Using the time evolution of the state of atom-field described by:
  \begin{eqnarray}\label{schrodinger equation}
      i\frac{\partial}{\partial t}|\Psi_{s}^{(1)}(t)\rangle=
      \mathbf{\mathcal{H}}_{eff} |\Psi_{s}^{(1)}(t)\rangle,
  \end{eqnarray}
       with $\mathbf{\mathcal{H}}_{eff} $ in (\ref{tamime effective Hamiltonian,rewrite})
       and the initial state in (\ref{first atom+field at t=0}),
       the state of the atom-field system after time $t$ is given by:
 \begin{eqnarray}\label{first atom+field at t}
      |\Psi_{s}^{(1)}(t)\rangle&= &\frac{1}{2 \sqrt{1+|\varepsilon_{1}|^2}}
      \sum_{n=0} ^{\infty} C_{n} \{ [(1+\varepsilon_{1}) e^{2i\lambda n f^2(n)t}-
      (1-\varepsilon_{1})]  |g,n\rangle\nonumber \\
      &+& [(1+\varepsilon_{1}) e^{2i\lambda n f^2(n)t}+(1-\varepsilon_{1})]|e,n\rangle\ \}.
 \end{eqnarray}
       The atomic velocities can be controlled such that every atom interacts for
       a fixed time $\tau$, with the radiation field inside the cavity. Suppose
       that the atom which is exiting the cavity is detected in the state $|e\rangle$,
       so the cavity field certainly collapses to
 \begin{eqnarray}\label{(first atom)field at tau}
      |\Psi_{F}^{(1)}(\tau)\rangle= \mathbf{\mathcal{N}}_{1}\sum_{n=0} ^{\infty}C_{n}
        \left [(1+\varepsilon_{1}) e^{2i\lambda n f^2(n)\tau}+(1-\varepsilon_{1})\right]  |n\rangle,
 \end{eqnarray}
      where $\mathbf{\mathcal{N}}_{1}$ is an appropriate normalization
      factor that  may be determined. Inserting $C_{n}$ from (\ref{bast nonlin}) in
      (\ref{(first atom)field at tau}) one can rewrite the field state as
  \begin{eqnarray}\label{vazeh}
      |\Psi_{F}^{(1)}(\tau)\rangle= \mathbf{\mathcal{N}}_{1} &\{ (1+\varepsilon_{1})&
       \mathbf{\mathcal{N}}(|z|^2) \sum_{n=0} ^{\infty}   \frac{z^n}{\sqrt{n!}\;
       [f(n)]!}\;e^{2i\lambda n f^2(n)\tau}|n\rangle\nonumber\\
      +&(1-\varepsilon_{1})& \mathbf{\mathcal{N}}(|z|^2) \sum_{n=0} ^{\infty}
       \frac{z^n}{\sqrt{n!}\; [f(n)]!}\;|n\rangle\}.
  \end{eqnarray}
      It is worth mentioning the fact that the normalization coefficient
      $\mathbf{\mathcal{N}}(|z|^2)$ for both nonlinear CS and GKCS is exactly the same.
      Noticing that $n f^2(n)=e_{n}$, at last the state in (\ref{vazeh}) can be written as
  \begin{eqnarray}\label{(first atom)field at tau, temporall+nonlinear}
      |\Psi_{F}^{(1)}(\tau)\rangle
      &=& \mathbf{\mathcal{N}}_{1} [(1+\varepsilon_{1}) |z,\alpha _{1}\rangle + (1-\varepsilon_{1})|z,f\rangle],
  \end{eqnarray}
      where $|z,\alpha _{1}\rangle$ is a GKCS introduced in (\ref {final bast GK z,alpha})
      with $\alpha _{1}\equiv-2\lambda \tau$. It is clear from
      (\ref{(first atom)field at tau, temporall+nonlinear}) that we have arrived at a
      superposition of GKCS and initial nonlinear CS. Now, if one prepares the atom
      in the initial state (\ref{ first atom}) with $\varepsilon_{1}=1$, namely the
      probabilities of the presence of the atom in $|e\rangle$ and $|g\rangle$ are equal,
      the cavity field will collapse to $|z,\alpha_{1}\rangle$. So, we have successfully
      generated a temporal stable state from initial nonlinear CS $|z,f\rangle$ that is not temporally stable.

      Now, assume that the second atom is injected into the cavity, while it is initially in the state described by
  \begin{eqnarray}\label{ second atom}
       |\Psi_{a}^{(2)}(0)\rangle=\frac{1}{\sqrt{1+|\varepsilon_{2}|^2}} (|e\rangle+\varepsilon_{2}|g\rangle).
  \end{eqnarray}
      If the second atom also interacts for a time $\tau$, with the cavity field which is
      described by (\ref {(first atom)field at tau}) and then, it is detected in the state
       $|e\rangle$, the cavity field collapses to
  \begin{eqnarray}\label{(second atom)field at tau}
      |\Psi_{F}^{(2)}(\tau)\rangle= &\mathbf{\mathcal{N}}_{2} &\sum_{n=0} ^{\infty} C_{n}
       \{ (1+\varepsilon_{1}) (1+\varepsilon_{2})e^{4i\lambda n f^2(n)\tau} \nonumber \\
      &+&2(1- \varepsilon_{1}\varepsilon_{2}) e^{2i\lambda n f^2(n)\tau} +
      (1-\varepsilon_{1})(1-\varepsilon_{2})  \}|n\rangle\ ,
  \end{eqnarray}
      where again $\mathbf{\mathcal{N}}_{2}$ is a normalization constant.
      Similar to the procedures we followed in
      (\ref{(first atom)field at tau})-(\ref{(first atom)field at tau,
      temporall+nonlinear}), the state in (\ref {(second atom)field at tau}) can be re-written as
  \begin{eqnarray}\label{(second atom)field temporral+nonlinear}
       |\Psi_{F}^{(2)}(\tau)\rangle
              &=&\mathbf{\mathcal{N}}_{2} [(1+\varepsilon_{1})
               (1+\varepsilon_{2}) |z,\alpha_{2}\rangle+2(1-
                \varepsilon_{1}\varepsilon_{2})|z,\alpha_{1}\rangle \nonumber\\
       &+&(1-\varepsilon_{1})(1-\varepsilon_{2}) |z,f\rangle],
  \end{eqnarray}
        where we have set $\alpha_{2}\equiv-4\lambda \tau$. Now,
        once again when the first and second atoms are initially prepared
        in $\frac{1}{\sqrt{2}}(|e\rangle + |g\rangle)$, i.e.,
        $\varepsilon_{1},\varepsilon_{2}=1$, the cavity field
        would be in $|z,\alpha_{2}\rangle$, i.e., in the GKCS.

       By further iteration of the above procedure it is easy to
       show that if the $m$th atom is initially in the state
  \begin{eqnarray}\label{ mth atom}
       |\Psi_{a}^{(m)}(0)\rangle=\frac{1}{\sqrt{1+|\varepsilon_{m}|^2}} (|e\rangle+\varepsilon_{m}|g\rangle),
 \end{eqnarray}
         and we enter $N$ atoms into the cavity, one by one, we can
         generate a superposition of $N$ temporally stable GKCSs and
         the initial state of the cavity field $|z,f\rangle$ that is
         not temporally stable. When all of $N$ atoms that are exiting
         one by one from the cavity are detected in the state $|e\rangle$, the state of the cavity field is given by
  \begin{eqnarray}\label{(Nth atom)field temporall}
       |\Psi_{F}^{(N)}(\tau)\rangle = \mathbf{\mathcal{N}}_{N} \left [\sum_{m=1} ^{N}
       C_{m}(\varepsilon_{1},\varepsilon_{2},...,\varepsilon_{N})|z,\alpha_{m}\rangle+C_{0}
       (\varepsilon_{1},\varepsilon_{2},...,\varepsilon_{N})|z,f\rangle \right],
  \end{eqnarray}
          where $\mathbf{\mathcal{N}}_{N}$ is an appropriate normalization constant and
          $\alpha_{m}\equiv-2^m \lambda \tau $.
          Again, if $\varepsilon_{1},\varepsilon_{2},$ $..., \varepsilon_{N}=1$, then $C_{0}=0$
          and $\{C_{m}\}_{m=1}^{N-1}=0$. As a result, the cavity field would be in
          $|z,\alpha_{N}\rangle$ which is exactly in the GKCSs family.

          We end this section with noticing that, the initial state of the cavity, i.e.,
          nonlinear CS may be generated experimentally. As an evidence recall that
          the generation of photon-added coherent states as a well-known class of
          nonlinear CSs reported by Zavatta et al in \cite{Zavatta}. Also, a theoretical
          scheme for generation of any class of nonlinear CSs in a micromaser
          using the intensity-dependent Jaynes-Cummings model was introduced in \cite{Naderi}.

\section{Summary and concluding remarks}
      In summary, using the intensity-dependent degenerate Raman interaction procedure has been introduced in the present paper
      and assuming the cavity field to be initially in the nonlinear CS (which
      is not temporally stable), the atoms that are in a superposition of the states
      $|g\rangle$ and $|e\rangle$ are sent inside the cavity, one by one. The atomic
      velocity can be controlled such that every atom interacts with the cavity field
      for a specific time interval $\tau$. If each of the atoms, which is exiting from
      the cavity is detected in the state $|e\rangle$, the cavity field collapses to a
      superposition of GKCSs and initial nonlinear CS. If $N$ atoms are sent inside the
      cavity, superposition of $N$ GKCSs and the initial state of the cavity will be generated.
      Preparing the initial state of the atom, such that the probabilities of being every atom
      in states $|g\rangle$ and $|e\rangle$ are equal, then the generated states reduce to a GKCS.
      In fact, we have generated temporally stable GKCS from a nonlinear CS which does not have
      this property. In addition to these, we have found  a more deep insight to the physical
      foundation of GKCSs, where we observed that according to our proposal,
      the parameter $\alpha$ in the phase factor
      of GKCSs in (\ref {final bast GK z,alpha}) explicitly depends on $\lambda$
     (the effective coupling constant of the atom-field)
      and $\tau$ (the time of the interaction between atom and
      field). Altogether, our proposal may be considered as the
      first step in producing the GKCSs. We hope that the proposed theoretical
      scheme will stimulate further better approaches to study the
      GKCSs, experimentally.

\vspace {2 cm}

   {\bf Acknowledgement:}
     We gratefully thank the referees for their comments which improved and clarified the context of the paper, considerably.

 \vspace {1.5 cm}

 {\bf FIGURE CAPTIONS}

   {\bf FIG. 1} Schematic diagram of degenerate $\Lambda$-type three-level atom
   interaction with the single-mode cavity field.

 \end{document}